\documentstyle[prb,aps,twoside,floats,epsf]{revtex}
\begin{document}
\draft
\twocolumn[\hsize\textwidth\columnwidth\hsize\csname
@twocolumnfalse\endcsname
\draft 
\title{
Assisted Tunneling in Ferromagnetic Junctions and 
 Half-Metallic Oxides
}
\author{   A.M. Bratkovsky  }
\address{
Hewlett-Packard Laboratories, 3500~Deer~Creek 
Road, Palo Alto, California 94304-1392 }
\date{Received January 29, 1998}

\maketitle
\begin{abstract}
Different mechanisms of spin-dependent tunneling are analyzed
with respect to their role in tunnel
magnetoresistance (TMR). Microscopic calculation within a realistic
model shows that direct tunneling in
iron group systems leads to about a 30\% change in resistance, 
which is close but lower than experimentally observed values. 
The larger observed values of the tunnel magnetoresistance
(TMR) might be a result of tunneling involving 
surface polarized states.
It is found that tunneling via resonant defect states in the
barrier radically decreases the TMR by order of magnitude.
One-magnon emission is shown to
reduce the TMR, whereas phonons increase the effect. 
The inclusion of both magnons and phonons reasonably explains an unusual 
bias dependence of the TMR.
The model presented here is applied qualitatively 
to half-metallics with 100\%  spin polarization, where 
one-magnon processes are suppressed and
the change in resistance in the absence of
spin-mixing on impurities may be arbitrarily large. Even in the 
case of imperfect
magnetic configurations, the resistance change can be a few 1000 percent.
Examples of half-metallic systems are  CrO$_2$/TiO$_2$ and 
CrO$_2$/RuO$_2$.

\end{abstract}

\vskip 2pc ] 


Tunnel magnetoresistance (TMR) in ferromagnetic junctions, 
first observed more than a decade ago,\cite{jul,maekawa} is of
fundamental interest and potentially applicable to
 magnetic sensors and  memory devices.\cite{tedrow}
This became particularly relevant after it was found
that the TMR for 3$d$ magnetic electrodes reached large
values at room temperature \cite{moodera},
and junctions demonstrated a non-volatile memory effect.

A simple model for spin tunneling, that has been formulated by Julliere
\cite{jul} and further developed in,\cite{stearns}
is expected to 
work rather well for iron, cobalt, and nickel based metals.
The microscopic model\cite{stearns} is in good agreement with
experimental results on bulk polarizations,\cite{tedrow,moodera}
and measured and calculated Fermi surfaces of $3d$ metals.
However, it disregards important points such as
impurity-assisted and inelastic scattering, tunneling into surface
states, and a reduced effective mass of carriers inside 
the barrier.
 These effects are important for proper understanding of the behavior of
actual devices, like peculiarities in their $I-V$ 
curves,\cite{amb_tunn1} and will be analyzed below.
A couple of {\em half-metallic} systems, which could in principle achieve
an ultimate magnetoresistance at room temperatures and low fields,
will be discussed.


The model that we will consider below includes a Hamiltonian 
${\cal H}_0$ for
non-interacting conducting spin-split electrons 
separated by a barrier,
electron-phonon interaction ${\cal H}_{ep}$, and exchange interaction 
of carriers with 
localized $d_l$ electrons ${\cal H}_{x}$, the latter giving rise
to an electron-magnon interaction. Impurity term ${\cal H}_i$ 
will correspond to a short-range confining potential producing 
defect states in the barrier,
\begin{equation}
{\cal H} = {\cal H}_0 + {\cal H}_{ep} + {\cal H}_x 
+ {\cal H}_i.
\end{equation}
Tunneling current is then evaluated within a general linear
response formalism.\cite{mahan}

Magnetoresistance (MR) is a relative change in a junction
conductance with respect to the change of mutual orientation of spins
from parallel ($G^{\rm P}$)  to antiparallel ($G^{\rm AP}$).
The MR depends only on effective polarization $P_{\rm FB}$
of tunneling electrons\cite{jul,amb_tunn1}
\begin{equation}
 {\rm MR} = {G^{\rm P}-G^{\rm AP} \over{ G^{\rm AP} }}=
{2P_{\rm FB}P'_{\rm FB}  \over { 1-P_{\rm FB}P'_{\rm FB}}}.
\label{eq:MR}
\end{equation}
The most striking feature of Eq.~(\ref{eq:MR}) is that the ${\rm MR}$
tends to infinity
when both electrodes are made of a 100\%
spin-polarized material ($P=P'=1$), because of a gap in the density of
states (DOS) for minority
carriers.
Such half-metallic behavior is rare,
but some materials possess this amazing property, most interestingly the
oxides CrO$_2$ and Fe$_3$O$_4$.\cite{kats} These oxides have potential
for future applications in combination with lattice-matching materials, as
illustrated below.

The full calculation of a TMR within microscopic model\cite{stearns}
due to direct tunneling can be performed numerically,\cite{amb_tunn1}
and it gives a value of  about 30\% at low biases. Note that 
electric field present in a biased barrier skews its shape, 
thus making it more transparent for `hot' electrons tunneling
at energies where the difference between the DOS of majority and
minority carriers is reduced. As a result the TMR in the direct tunneling
decreases with the increased bias. 

In a half-metallic case  
we obtain {\em zero} conductance $G^{AP}$ in the AP 
configuration at biases within the half-metallic band gap.\cite{amb_tunn1}
Even at 20$^\circ$ deviation from the AP configuration, the value of MR 
 exceeds 3,000\% within the half-metallic gap, 
and this is indeed a very large value.\cite{amb_tunn1}


Presence of impurity/defect states in the barrier would result in
a resonant tunneling of electrons.
Comparing the direct and the impurity-assisted contributions to
conductance, it is easy to
see that the latter dominates when the density of impurity states
exceeds $\sim 10^{17}$cm$^{-3}$eV$^{-1}$.\cite{amb_tunn1} 
When impurities take over, 
the magnetoresistance is again given by the Julliere's
formula where the effective polarization $\Pi$ is that of impurity 
`channels',\cite{amb_tunn1}
\begin{equation}
{\rm MR}_1=2\Pi~\Pi'/(1-\Pi~\Pi'),
\label{eq:mr1}
\end{equation}
and it gives a value of about 4\% for MR$_1$ in the 
case of Fe with non-magnetic impurities.
The value of MR is reduced
since generally $\Pi<P_{\rm FB}$ because of mixing of the tunneling 
electron wave function with non-polarized defect states.
In the case of magnetic impurities (spin-flip centers) 
the TMR will be even smaller. At the same time, conductance may be
substantially increased. These predictions\cite{amb_tunn1} 
have been confirmed 
experimentally.\cite{janice,mood_def} Resonant diode type of
structure gives similar results.\cite{amb_tunn1}


Direct tunneling, as we have seen, gives TMR of about 30\%,
whereas in recent experiments TMR is well above this value,
approaching 40\%.\cite{janice,mood_large}
It would become clear below that this enhancement is unlikely to come
from the inelastic processes. 

Up to now we have disregarded
the possibility of localized states at metal-oxide interfaces.
Keeping in mind that the usual barrier AlO$_x$ is amorphous,
the density of such states may be high.
We have to take into account tunneling into/from those states.
The corresponding tunneling MR is found to be 
\begin{eqnarray}
{G_{\rm bs}(\theta)\over A} &=& {e^2\over \pi\hbar} B\overline{D}_s(1+P_{\rm
FB}P_s\cos(\theta)),\nonumber\\
P_s &=& {D_{s\uparrow} - D_{s\downarrow}\over{D_{s\uparrow} 
+ D_{s\downarrow}}},\hspace{.2in}
\overline{D}_s = \frac{1}{2}
(D_{s\uparrow}+D_{s\downarrow}),
\label{eq:Gsurface}
\end{eqnarray}
where $P_s$ is the polarization and $\overline{D}_s$
is the average density of surface states, and $\theta$ 
is the mutual angle between moments on electrodes.
The parameter 
$B \sim [2\pi\hbar^2m\kappa/(m_2^2w)]\exp(-2\kappa w)$, where
$w$ is the barrier width, $\kappa$ is the
absolute value of electron momentum under the barrier,
$m$ and $m_2$ are the free electron mass and the effective mass in
the barrier, respectively.
The corresponding magnetoresistance would be 
${\rm MR_{bs}} = 2P_{\rm FB}P_s/(1-P_{\rm FB}P_s)$.

It is easy to show that the bulk-to-surface
conductance exceeds the bulk-to-bulk one at 
densities of surface states $D_s>D_{sc}\sim 10^{13}$cm$^{-2}$eV$^{-1}$
per spin, comparable to those found at some metal-semiconductor 
interfaces.

If on both sides of the barrier
the density of surface states is above the critical value
$D_{sc}$,
the magnetoresistance would be due to surface-to-surface  tunneling
with a value given by
$ {\rm MR_{\rm ss}} = 2P_{s1}P_{s2}/(1-P_{s1}P_{s2}). $
If the polarization of surface states is larger than that of the bulk,
as is often the case even for imperfect surfaces,\cite{cab}
then it would result in enhanced TMR.

\begin{figure}[t]
\epsfxsize=2.2in
\epsffile{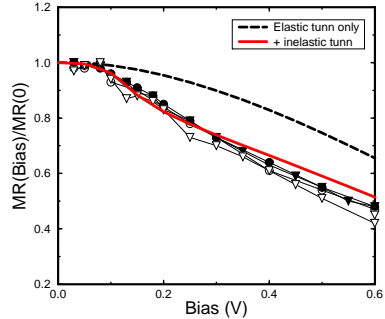}
\caption{
Fit to experimental data for the magnetoresistance
of CoFe/Al$_2$O$_3$/NiFe tunnel junctions [9] 
with inclusion of elastic and inelastic (magnons and phonons)
tunneling. 
The fit gives for magnon DOS $\propto \omega^{0.65}$ which is 
close to a standard bulk spectrum  $\propto \omega^{1/2}$.$^{13}$
\label{fig:fit}
}
\end{figure}

Inelastic processes with excitation of magnon or phonon modes 
introduce new energy scales into the problem (30-100 meV) which
correspond to a region where unusual $I-V$ tunnel characteristics are
seen (Fig.~\ref{fig:fit}).
We obtain for magnon-assisted 
inelastic tunneling current at $T=0$:
\begin{eqnarray}
I^x_{\rm P} &=& {2\pi e\over \hbar}\sum_\alpha X^\alpha g^L_\downarrow g^R_\uparrow
\int d\omega 
\rho^{mag}_{\alpha}(\omega) (eV-\omega)\theta(eV-\omega),\nonumber\\
I^x_{\rm AP} &=& {2\pi e\over \hbar}\biggl[
X^R g^L_\uparrow g^R_\uparrow
\int d\omega \rho^{mag}_{R}(\omega) (eV-\omega)\theta(eV-\omega)\nonumber\\
&+&
X^L g^L_\downarrow g^R_\downarrow
\int d\omega \rho^{mag}_{L}(\omega) (eV-\omega)\theta(eV-\omega)\biggr],
\label{eq:Ix0}
\end{eqnarray}
where $X$ is the incoherent tunnel exchange vertex,
$\rho_\alpha^{mag}(\omega)$ is the magnon density of states
that has a
general form $\rho_\alpha^{mag}(\omega)
=(\nu+1)\omega^{\nu}/\omega_0^{\nu+1}$, the exponent
$\nu$ depends on a type of spectrum, 
$\omega_0$ is the maximum magnon frequency, $g^{L(R)}$ marks
the corresponding electron density of states on left (right)
electrode,  $\theta(x)$ is the step function, $\alpha=L, R$.

For phonon-assisted current at $T=0$ we obtain 
\begin{eqnarray}
I^{ph}_{\rm P} &=& {2\pi e\over \hbar}\sum_{a\alpha} g^L_a g^R_a
\int d\omega  \rho^{ph}_\alpha(\omega) 
P^\alpha_\omega(eV-\omega)\theta(eV-\omega),\nonumber\\
I^{ph}_{\rm AP} &=& {2\pi e\over \hbar}\sum_{a\alpha} g^L_a g^R_{-a}
\int d\omega  \rho^{ph}_\alpha(\omega) 
P^\alpha_\omega(eV-\omega)\theta(eV-\omega),
\label{eq:Ip0}
\end{eqnarray}
where $P_\omega$ is the phonon vertex,
$P_\omega/X = \gamma \omega/\omega_D$, where $\gamma$ is the constant
and $\omega_D$ is the Debye frequency, $a$ is the spin index, 
$\rho_\alpha^{ph}$ is the phonon density of states,
$\alpha$ marks electrodes and the barrier.

The elastic and inelastic contributions together will define the total
junction conductance $G=G(V,T)$ as a function of the bias $V$ and
temperature $T$.
We find that the inelastic contributions from magnons and phonons
(\ref{eq:Ix0})-(\ref{eq:Ip0}), respectively, grow as 
$G^x(V,0) \propto (|eV|/\omega_0)^{\nu+1}$ and 
$G^{ph}(V,0) \propto (eV/\omega_D)^4$
at low biases. 
These contributions saturate at higher biases: $G^x(V,0) \propto
1-\frac{\nu+1}{\nu+2}\frac{\omega_0}{|eV|}$ at $|eV|>\omega_0$;
$G^{ph}(V,0) \propto 
1-\frac{4}{5}\frac{\omega_D}{|eV|}$ at $|eV|>\omega_D$.
This behavior would lead to sharp features in the
$I-V$ curves on a scale of 30-100 mV (Fig.~\ref{fig:fit}).

It is important to highlight the opposite effects of phonons and magnons
on the TMR. If we take the case of the same electrode materials and
denote $D=g_\uparrow$ and $d=g_\downarrow$ then we see that
$G^x_{\rm P}(V,0) - G^x_{\rm AP}(V,0) 
\propto - (D-d)^2(|eV|/\omega_0)^{\nu+1}<0$, whereas
$G^{ph}_{\rm P}(V,0) - G^{ph}_{\rm AP}(V,0) 
\propto +(D-d)^2(eV/\omega_D)^4>0 $, i.e.
spin-mixing due to magnons decreases the TMR, whereas the phonons tend to
reduce the negative effect of magnon emission.\cite{zhang} 
Different bias and temperature dependence can make possible a separation 
of these two contributions, which are of opposite sign.

At finite temperatures we obtain the contributions
of the same respective sign as above.
For magnons: 
$G^x_{\rm P}(0,T) - G^x_{\rm AP}(0,T) 
\propto - (D-d)^2 (-TdM/dT)<0$, 
where $M=M(T)$ is the magnetic moment of the electrode at a given 
temperature $T$. The phonon contribution is given by a standard 
Debye integral with the following results:
$G^{ph}_{\rm P}(0,T) - G^{ph}_{\rm AP}(0,T) 
\propto +(D-d)^2(T/\omega_D)^4>0 $ at $T\ll\omega_D$, and 
$G^{ph}_{\rm P}(0,T) - G^{ph}_{\rm AP}(0,T) \propto
+(D-d)^2(T/\omega_D)$
 at $T\gtrsim \omega_D$. 
It is worth mentioning that the magnon excitations 
are usually cut off by e.g. the anisotropy
energy $K_{\rm an}$ at some $\omega_c$. Therefore, at
low temperatures  the conductance 
at small biases will be almost constant.


It is very important that in the case of {\em half-metallics}
$P_{\rm FB}=\Pi=1$,
and even with an imperfect barrier the magnetoresistance can, 
at least in principle, reach any value
limited by only spin-flip processes in the barrier/interface.
We should note that the {\em one}-magnon excitations in
half-metallics are suppressed by the half-metallic gap in 
electron spectrum.
Spin-mixing can only occur on magnetic impurities
in the barrier or interface, the allowed {\em two}-magnon
excitations in the electrodes do not result in spin-mixing.

\begin{figure}[t]
\epsfxsize=2.4in
\epsffile{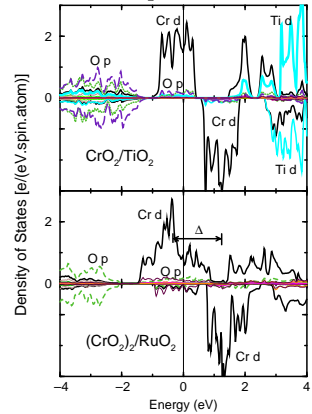}
\caption{
Density of states of
CrO$_2$/TiO$_2$ (top panel)
and (CrO$_2$)$_2$/RuO$_2$ (bottom panel) half-metallic 
layered structures calculated with the
use of the LMTO method.
\label{fig:cro2}
}
\end{figure}

The examples of systems 
with half-metallic behavior are CrO$_2$/TiO$_2$ and 
CrO$_2$/RuO$_2$ (Fig.~\ref{fig:cro2}).\cite{amb_tunn1}
They are based on half-metallic CrO$_2$, and all materials 
have the rutile structure with almost perfect lattice 
matching. This should yield a good interface
and help in keeping the system at desired stoichiometry.
TiO$_2$ (RuO$_2$) are used as the barrier (spacer) oxides.
The half-metallic behavior of the corresponding 
multilayer systems is  demonstrated by the
band structures calculated within the linear muffin-tin orbitals method (LMTO)
in a supercell geometry with [001] growth
direction (Fig.~\ref{fig:cro2}).
The calculations show that CrO$_2$/TiO$_2$ is a perfect half-metallic, whereas
(CrO$_2$)$_2$/RuO$_2$ is a weak half-metallic, since 
there is some small DOS around $E_F$.

The main concerns for achieving a very
large value of magnetoresistance with half-metallics
will be spin-flip centers
and imperfect alignment of moments, provided that other,
e.g. many-body, effects are small.
As for conventional tunnel junctions, the
present results show that the defect states in the barrier, 
or a resonant state like in a resonant tunnel diode type of
structure, reduces their magnetoresistance by several times but may
dramatically increase the current through the structure.
The inelastic processes are  responsible for the unusual shape of
the $I-V$ curves at low biases, and their temperature behavior,
which is also affected by impurity-assisted tunneling.
The surface states assisted tunneling may lead to enhanced TMR,
if their polarization is higher than that of the bulk.
This could open up ways to improving performance of 
ferromagnetic tunnel junctions.

I am grateful to T.~Anthony, G.A.D.~Briggs, J. Brug,
J. Moodera, N. Moll, J.~Nickel, and R.S.~Williams for 
useful discussions.

\end{document}